\documentclass[twocolumn]{aastex63}
\shorttitle{UV Transit Features}
\shortauthors{Lothringer et al.}

\usepackage{amsmath}
\usepackage{xspace}
\usepackage{float}

\newcommand{\microns}{$\mu$m}

\begin{document}
\title{UV Exoplanet Transmission Spectral Features as Probes of Metals and Rainout}
\author[0000-0003-3667-8633]{Joshua D. Lothringer}
\affiliation{Department of Physics and Astronomy, Johns Hopkins University, Baltimore, MD, USA}

\author{Guangwei Fu}
\affiliation{Department of Astronomy, University of Maryland, College Park, MD, USA}

\author[0000-0001-6050-7645]{David K. Sing}
\affiliation{Department of Physics and Astronomy, Johns Hopkins University, Baltimore, MD, USA}
\affiliation{Department of Earth \& Planetary Sciences, Johns Hopkins University, Baltimore, MD, USA}

\author[0000-0002-7129-3002]{Travis S. Barman}
\affiliation{Lunar and Planetary Laboratory, University of Arizona, Tucson, AZ, USA}

\vspace{0.5\baselineskip}
\date{\today}
\email{jlothri1@jhu.edu}

\begin{abstract}
	The low-resolution transmission spectra of ultra-hot Jupiters observed shortward of 0.5~\microns{} indicate strong absorption at short-wavelengths. Previous explanations have included scattering, photochemistry, escaping metals, and disequilibrium chemistry. In this Letter, we show that slopes and features shortward of 0.5 \microns{} can be caused by opacity not commonly considered in atmosphere models of exoplanets but guaranteed to be present if conditions are near chemical equilibrium including Fe I, Fe II, Ti I, Ni I, Ca I, Ca II, and SiO. Even relatively trace species (e.g., Cr I and Mn I) can contribute through strong lines in the UV and blue-optical. Using the PHOENIX atmosphere model, we describe how the short-wavelength transit spectrum varies with equilibrium temperature between 1000~K and 4000~K, as well as the effect that the rainout of condensates has at these wavelengths. We define two spectral indices to quantify the strength of the NUV and blue absorption compared to that in the red-optical, finding that the NUV transit depth will significantly exceed the transit depth from Rayleigh scattering alone for all hot Jupiters down to around 1000~K. In the blue-optical, hot Jupiters warmer than 2000 K will have transit depths larger than that from Rayleigh scattering, but below 2000~K, Rayleigh scattering can dominate, if present. We further show that these spectral indices may be used to trace the effects of rainout. We then compare our simulated transit spectra to existing observations of WASP-12b, WASP-33b, WASP-76b, and WASP-121b.
	
\end{abstract}


\section{Introduction}


The spectral footprint of different atmospheric components can be identified through transmission spectroscopy. Light from the host star is effectively filtered through a planet's terminator, allowing us to identify and characterize the composition and, to a lesser degree, the temperature of the atmosphere.

By combining HST/STIS and WFC3 grisms, complete low-resolution transit spectra between 0.3 and 1.7 \microns{} have been obtained for several planets to date \citep[e.g.,][]{sing:2016}. In this range, one usually seeks to detect and characterize molecular absorption from H$_2$O, TiO, and VO, as well as broad alkali lines from Na and K. Aerosol opacity, from either condensate clouds or photochemical hazes, can be inferred by detecting a uniform gray opacity or a scattering slopes toward short wavelengths. 

Large transit depths at short wavelengths have been found in a number of exoplanets \citep[e.g.,][]{ballester:2007,pont:2008,etangs:2008,sing:2008b} and have frequently been well-fit by scattering, often approximated by 

\begin{equation}
	\sigma = \sigma_0(\lambda/\lambda_0)^{-\alpha},
\end{equation}
\noindent where $\lambda$ is the wavelength, $\lambda_0$ is the reference wavelength, $\sigma$ is the scattering cross-section, $\sigma_0$ is the cross-section at the reference wavelength, and $\alpha=4$ in the case of Rayleigh scattering. Unocculted starspots can also cause increased short-wavelength transit depths \citep{pont:2013,mccullough:2014,rackham:2018}.

Recent observational and theoretical studies into the hottest known Jovian exoplanets, called ultra-hot Jupiters, have shown a plethora of atomic species and their ions can exist in these planets' atmosphere \citep[e.g.,][]{lothringer:2018b,lothringer:2019,arcangeli:2018,parmentier:2018,hoeijmakers:2019,yan:2018,casasayas:2019}. At equilibrium temperatures of more than 2000 K, molecules begin to thermally dissociate, atoms ionize, and condensation of even the most refractory elements ceases. The effect of these atomic species on short-wavelength exoplanet transmission spectra has yet to be fully explored.

Indeed, transmission spectra of ultra-hot Jupiters WASP-12b, WASP-33b, WASP-76b, and WASP-121b all show significant absorption at wavelengths less than 0.5~\microns{}. The moderate slope in the optical transmission spectrum of WASP-12b has thus far been interpreted as being caused by scattering by hazes \citep{sing:2013,kreidberg:2015}, while the NUV spectrum from HST/COS shows evidence of escaping metals \citep{fossati:2010,haswell:2012}. The short-wavelength absorption in WASP-33b was interpreted as being from AlO, which would be highly out of equilibrium, but could explain absorption between 0.4 and 0.5 \microns{} \citep{vonessen:2018}. The slope in WASP-121b was interpreted as being from the photochemical product SH \citep[see also \citealt{zahnle:2009b}]{evans:2018}, though additional observations with \text{Swift}/UVOT and HST/STIS/E230M show that this slope continues below 0.3 \microns{} with evidence of escaping Fe II and Mg II \citep{salz:2019,sing:2019}. Fe I has also been detected in WASP-121b in multiple datasets \citep{gibson:2020,bourrier:2020,cabot:2020}.

The recent transmission spectrum of WASP-76b (T$_{eq}$=2160 K) also show large NUV and blue transit depths in addition to TiO and H$_2$O opacity \citep{fu:2020}. These observations were well-fit by a self-consistent atmosphere model at solar metallicity, without the need to appeal to disequilibrium processes. It was shown that the large transit depths at short wavelengths can be caused by opacity not commonly considered in atmosphere models of exoplanets, but guaranteed to be present in chemical equilibrium including Fe and SiO. The presence of Fe I in WASP-76b has also been detected with ground-based high-resolution observations, with evidence that it is gaseous on the evening terminator, but condenses on the nightside \citep{ehrenreich:2020}.

In this Letter, we study these important opacity sources further by computing additional hot Jupiter short-wavelength transmission spectra. In Section~\ref{sec:methods}, we describe our modeling setup. In Section~\ref{sec:results}, we discuss our model spectrum of WASP-76b, explore how the short-wavelength transit spectrum varies with temperature, investigate atmospheric heating by NUV opacity sources, and compare our models to observations of other ultra-hot Jupiters. In Section~\ref{sec:conclude}, we close with further discussion and conclusions.

\section{Methods}\label{sec:methods}

We use the PHOENIX atmosphere model to calculate the composition, structure, opacity, and transmission spectrum of several hot Jupiters. Our model setup is similar to previous studies of ultra-hot Jupiters \citep{lothringer:2018b,lothringer:2019}. PHOENIX self-consistently calculates the composition and structure of an atmosphere assuming chemical and radiative-convective equilibrium, including the irradiation from a primary companion \citep{hauschildt:1999,barman:2001}. We consider some models with rainout chemistry, where an element will be depleted in layers above if it is present in condensates in the lower atmosphere. Our rainout models assume efficient settling and no vertical mixing and thus represent the limiting behavior of rainout.

Thanks to its large EUV-to-FIR opacity database of atomic opacity up to uranium \citep{kurucz:1994a,kurucz:1994b,kurucz:1994c} and over 130 molecular species, PHOENIX is ideal to model ultra-hot exoplanets, particularly at short wavelengths. The dominant molecular opacity in our model are SiO \citep{kurucz:1993}, TiO \citep{schwenke:1998}, and H$_2$O \citep{barber:2006}.

\begin{figure*}[t]
	\centering
	\includegraphics[width=7in]{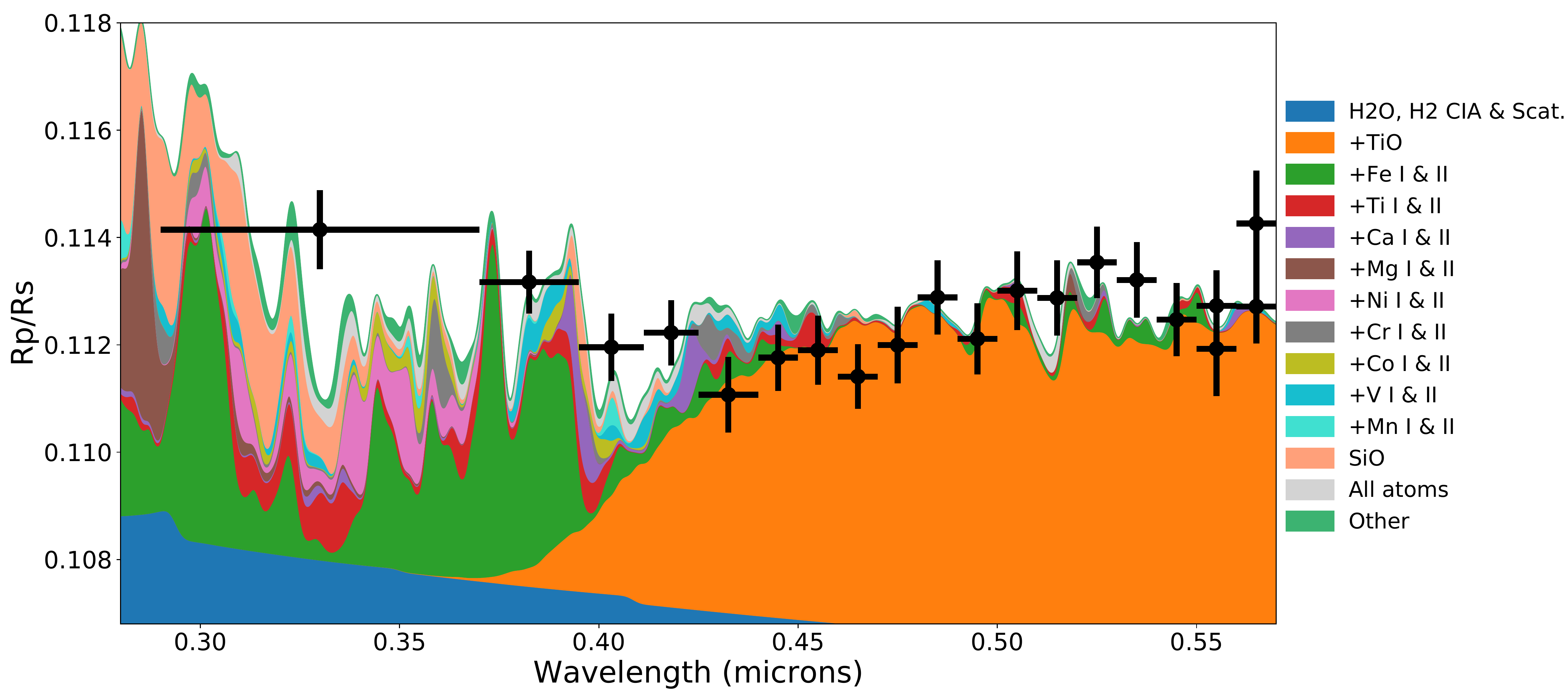} 
	\caption{Contribution of various opacity sources to transit spectrum of WASP-76b. \label{fig:contribution}}
\end{figure*}

\section{Results and Discussion}\label{sec:results}

\subsection{WASP-76b}

\cite{fu:2020} present a transit spectrum of WASP-76b (T$_{eq}$=2180~K) from 0.3-5 \microns{} using data from HST/STIS/G430L, HST/STIS/G750L, HST/WFC3/G141, and Spitzer channels 1 and 2. Retrievals using PLATON \citep{zhang:2019} and ATMO \citep{evans:2017} require either an unphysically strong scattering slope or a high Fe abundance, respectively, to fit the short-wavelength slope, however the spectrum is well-fit by a self-consistent cloud-free chemical-equilibrium solar-metallicity PHOENIX model. TiO and H$_2$O are evident in the spectra as strong absorption between 0.45-1.0 \microns{} and 1.3-1.6 \microns{}, respectively. Fe alone does not provide enough opacity to fit the slope towards increasing transit depth with decreasing wavelength shortward of 0.45 \microns{}. Additional opacity from a number of metals and molecules contribute to fit the observations, including Ti I, Ni I, Ca I \& II, and SiO.

Figure~\ref{fig:contribution} shows the contribution of these various atoms and molecules to the transit spectrum of WASP-76b compared to the data. As opacity from TiO begins to drop off shortward of 0.45 \microns{}, Fe opacity begins to increase. Strong lines from Ca and Cr add to a bump at 0.43 \microns{}. This region, between 0.43 and 0.5 \microns{}, does not agree well with the data, possibly indicating some of these species may not be present in the gas phase or are otherwise weaker than our model predicts. The Ca II H \& K lines are evident at 0.39 \microns{} with additional opacity from V I and a band of Fe I lines between 0.37 and 0.39 \microns{}. Strong Fe I and Cr I lines at 0.36 \microns{} produce another bump in transit depth. Ni I and II provide a forest of lines between 0.34 and 0.36 \microns{}. SiO becomes a major opacity source shortward of 0.35 \microns{} and helps explain the shortest-wavelength bin of the HST/STIS observations when combined with a major Fe I band at 0.3 \microns{}. While shortward of the current WASP-76b HST/STIS data,the Mg II doublet has been observed well past the planetary Roche lobe in WASP-12b \citep{fossati:2010} and WASP-121b \citep{sing:2019}, but not in the much cooler HD~209458b \citep{cubillos:2020}.

\begin{figure}
	\centering
	\includegraphics[width=3.5in]{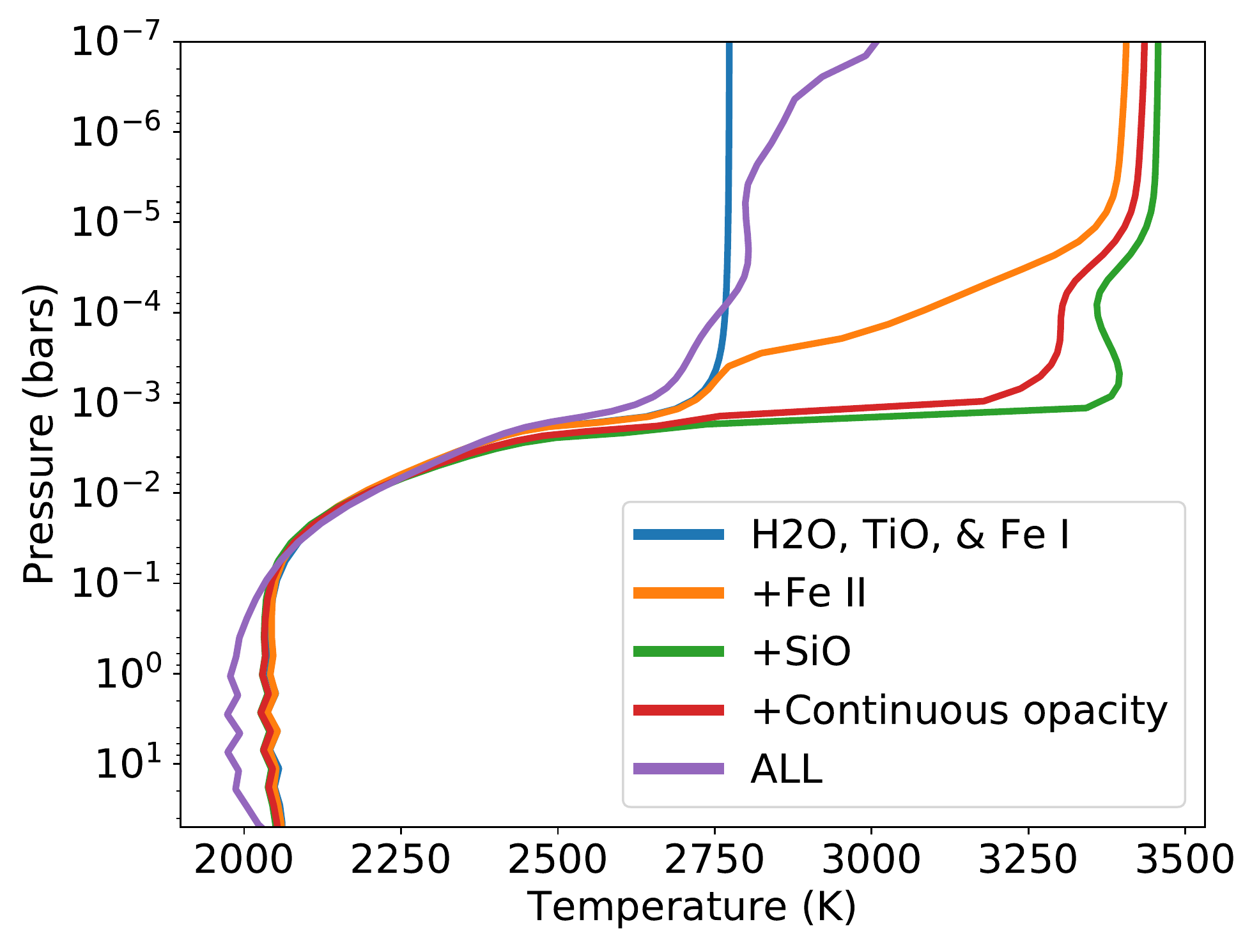}
	\caption{Temperature structures from models of WASP-76b with different opacity sources included. Continuous opacity includes photoionization cross-sections from atoms including H I, He I, C I, N I, O I, Na I, Mg I \& II, Al I, Si I \& II, S I, Ca I \& II, and Fe I, as well as H$^-$ and collision-induced-absorption.  \label{fig:heating}}
\end{figure}

Figure~\ref{fig:heating} shows the converged temperature structure of several models including different sets of UV opacity sources in order to examine their contribution to atmosphere heating. The large temperature inversion is caused by heating from the absorption of NUV and optical irradiation by atoms and molecules like TiO, Fe I \& II, and SiO, as has been shown for other ultra-hot Jupiters \citep{lothringer:2018b,gandhi:2019}. This level of atmospheric heating has also been suggested in the context of WASP-121b's UV absorber \citep{evans:2018}. However, the inclusion of all opacities results in some cooling of the upper atmosphere, presumably from radiative cooling from certain lines. While outside the scope of this Letter, the influence of trace metals on the energy balance in the upper atmosphere is worth future investigation.

\subsection{Generic Hot Jupiter}\label{sec:generic}

\begin{figure*}[h!]
	\centering
	\vspace{-16pt}
	\includegraphics[width=7in]{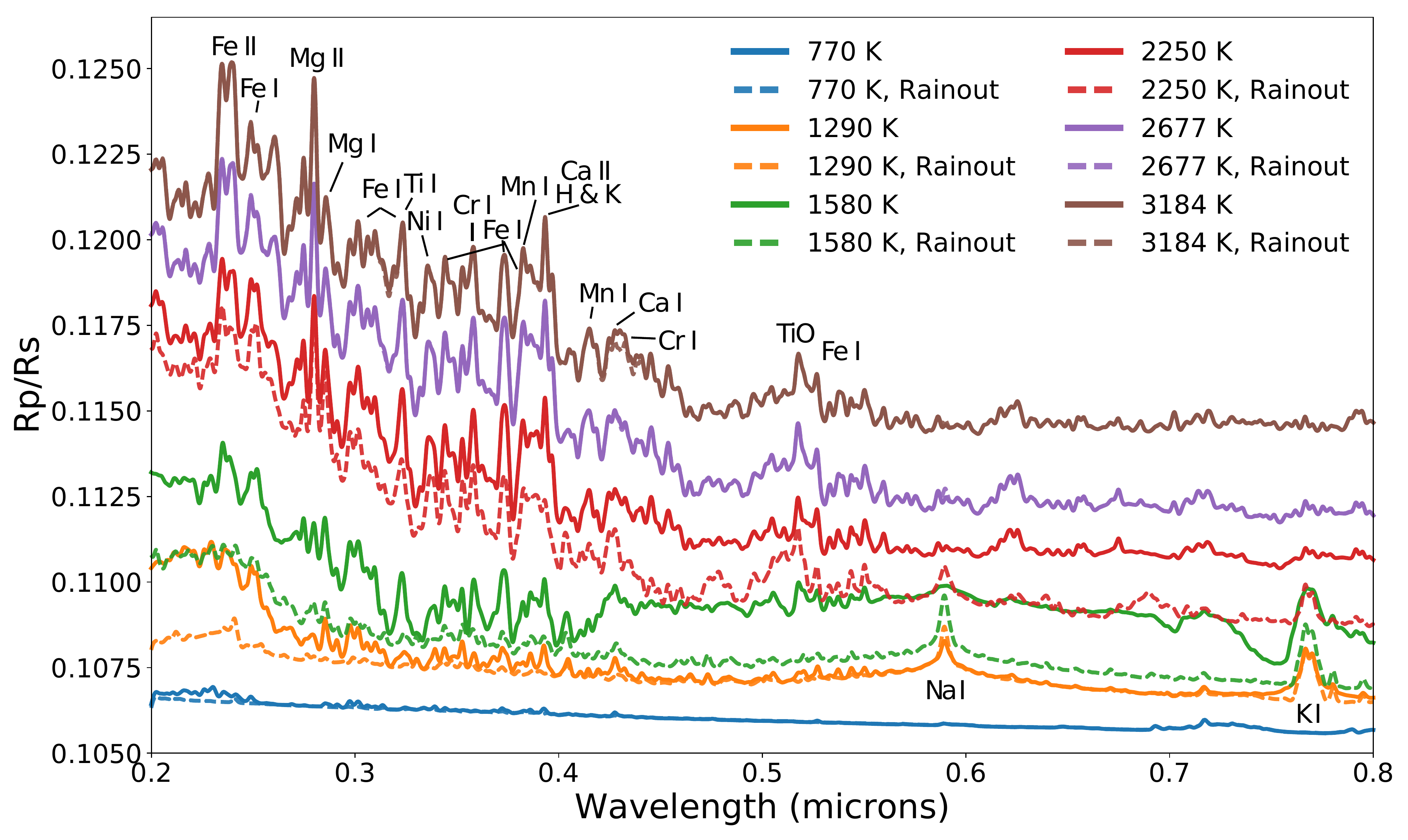} 
	\vspace{-10pt}
	\caption{Transit spectra between 0.2 and 0.8 \microns{} of the fiducial generic hot Jupiter at different temperatures. Solid lines indicate local chemical equilibrium, while dashed lines indicate models with rainout chemistry. For the 2677 K and 3184 K models, no condensation takes place so the dashed line is identical to the solid line. No offset has been applied to these models, rather their spacing is the natural effect of their different scale heights as each model has the same radius at the bottom of the model (i.e., 1.5~R$_J$ at $\tau_{1.2\mu{m}}=1000$). \label{fig:diffT}}
	\vspace{-30pt}
\end{figure*}

To explore how the short-wavelength transit spectrum varies with temperature, we calculated atmosphere models for a generic hot Jupiter at several equilibrium temperatures: 770, 1100, 1290, 1580, 1890, 2250, 2680, 3184, and 4500~K, corresponding to 8$\times$, 4$\times$, 3$\times$, 2$\times$, 1.4$\times$, 1$\times$, 0.7$\times$, 0.5$\times$, and 0.25$\times$ the orbital distance of the 2250~K model, respectively. The properties of the generic hot Jupiter are identical to those used in \cite{lothringer:2018b} and \cite{lothringer:2019}, namely 1 M$_J$ and 1.5 R$_J$ orbiting a 1.5 R$_{Sun}$ 7200 K F0 star. Figure~\ref{fig:diffT} shows a subset of these, including the 770, 1290, 1580, 2250, 2677, and 3184 K models. We also ran models with the same temperature structure, but with rainout included in the chemistry, indicated by dashed lines in Figure~\ref{fig:diffT}. 

Overall, the short-wavelength transit slope decreases in magnitude with decreasing temperature as metals and other species condense out of the gas phase. This is further enhanced in the rainout models where an element will be depleted in the upper atmosphere if it condenses in the lower atmosphere. By approximately 1300~K, the only significant short-wavelength opacity remaining in the model with rainout is the bound-free opacity from K. The effect of rainout on the NUV and blue-optical opacities was previously pointed out in the context of HD~209458b in \cite{barman:2007}. At red wavelengths, TiO and VO begin to rain out of the atmosphere around 2000 K, revealing the strong Na and K resonance lines, which themselves condense around 1000 K.

\subsubsection{Spectral Indices}

We quantify the strength of the short-wavelength opacity by defining two spectral indices. The first, $\Delta$R$_{p,NUV-Red}$, compares the transit radius of the planet between 0.2-0.3~\microns{} with the radius between 0.6-0.7~\microns{}. The indices are normalized by the atmospheric scale height at the equilibrium temperature (H=$kT_{eq}/\mu{g}$) such that a dimensionless quantity can be compared to across planets with varying gravity and temperatures. Differences in the surface gravity and temperature structure on planets at the same equilibrium temperature can still result in changes up to about 10\% in these indices. We also note that the NUV spectrum between 0.2-0.3~\microns{} may probe high enough in the atmosphere to be dominated by escaping gas \citep[e.g.,][]{vidal-madjar:2013,fossati:2010,sing:2019}, so the transit depths that we calculate in that range are likely a lower limit. The second spectral index, $\Delta$R$_{p,Blue-Red}$, compares the transit radius of the planet between 0.3-0.4~\microns{} (approximately the bluest bin in G430L observations) with the radius between 0.6-0.7~\microns{}. Formally,

\begin{equation}
	\Delta R_{p,NUV-Red} = \frac{R_{p,0.2-0.3\mu{m}}-R_{p,0.6-0.7 \mu{m}}}{H_{eq}}
\end{equation}

and

\begin{equation}
\Delta R_{p,Blue-Red} = \frac{R_{p,0.3-0.4\mu{m}}-R_{p,0.6-0.7 \mu{m}}}{H_{eq}}
\end{equation}

\begin{figure*}
	\centering
	\gridline{\fig{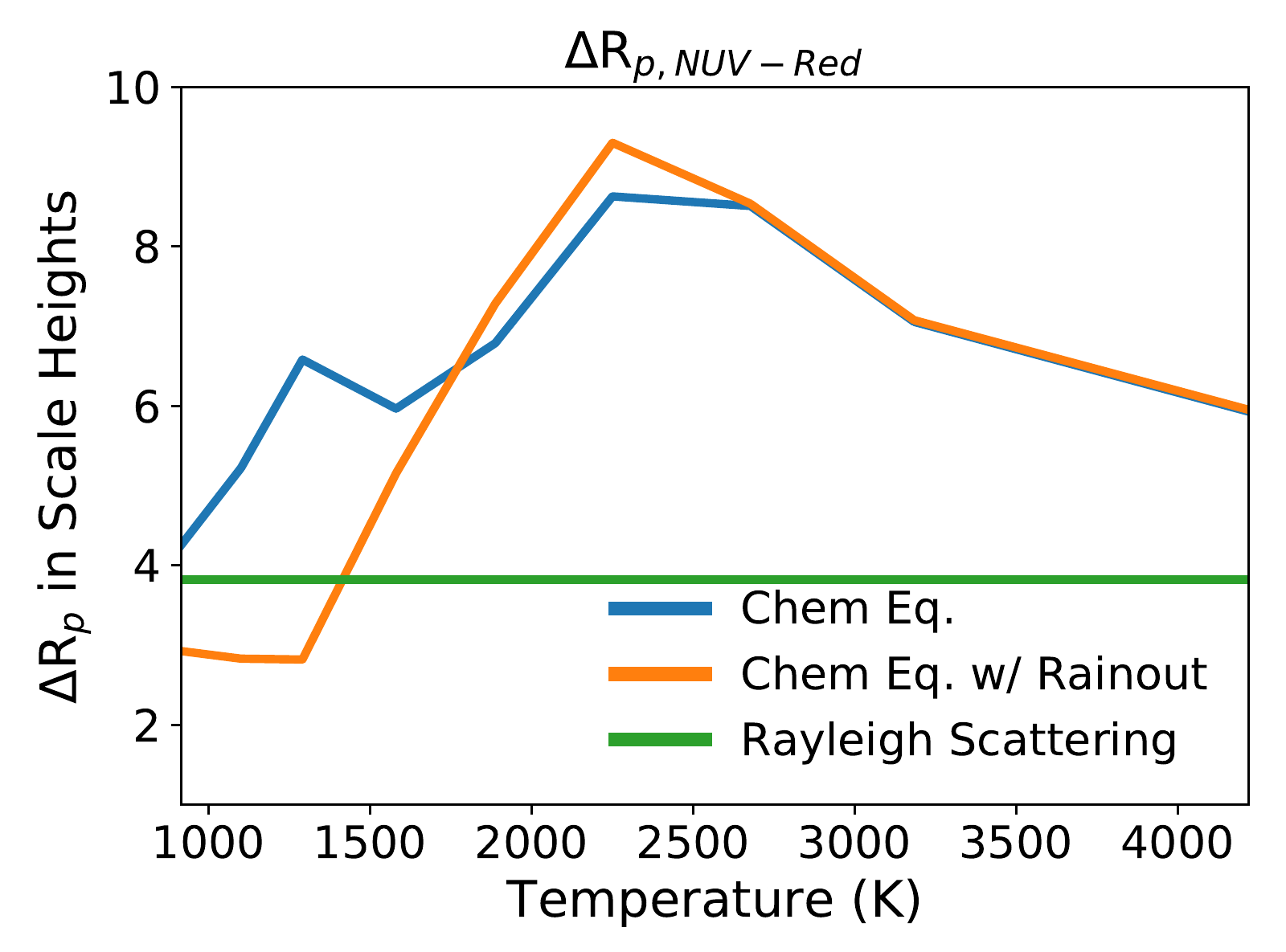}{0.5\textwidth}{}
		\fig{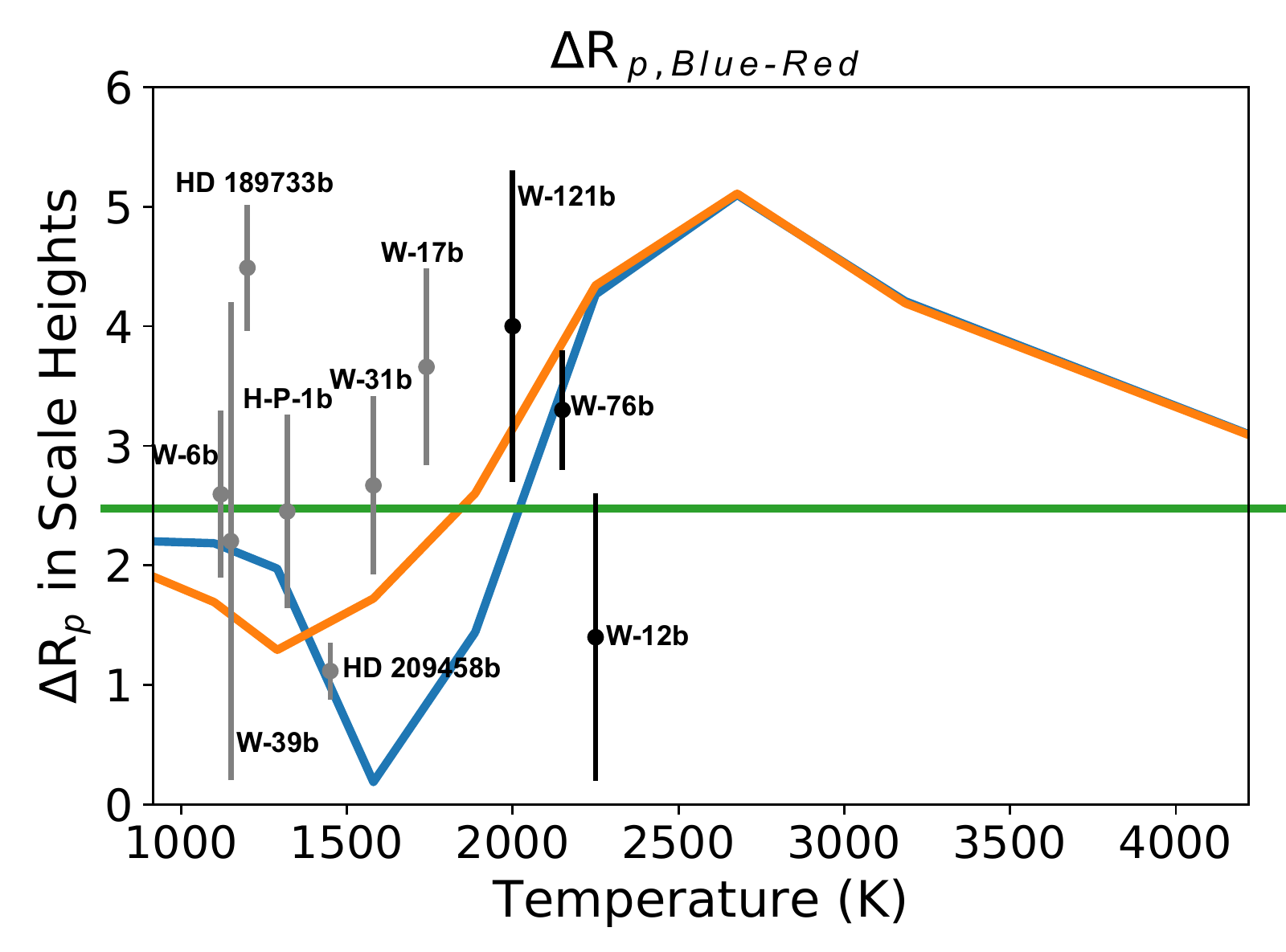}{0.5\textwidth}{}}
	\centering
	\caption{Left: The $\Delta$R$_{p,NUV-Red}$ spectral index, quantifying the difference between the transit radius, in terms of scale heights, between 0.2-0.3 \microns{} and 0.6-0.7 \microns{} as a function of temperature. Right: Same as left, but for the $\Delta$R$_{p,Blue-Red}$, measured between 0.3-0.4 \microns{} and 0.6-0.7 \microns{}. Also plotted is the expected slope from Rayleigh scattering.
		\label{fig:index}}
\end{figure*}

We plot these two indices as a function of equilibrium temperature in Figure~\ref{fig:index} using the models described in Section~\ref{sec:generic}. $\Delta$R$_{p,NUV-Red}$ remains high for all temperatures, and, in particular, is above the slope expected from Rayleigh scattering alone, assuming opacity of the form $\sigma=\sigma_0(\lambda/\lambda_0)^{-4}$, for all scenarios except the coldest rainout models. Up to 9 scale heights are probed across the spectral index, implying the spectral imprint of the opacity sources we discuss here should be readily observable for a wide range of systems. 

At high temperatures, models with chemical equilibrium are similar to those with rainout chemistry, however $\Delta$R$_{p,NUV-Red}$ begins to decrease more steeply for models with rainout compared to those in chemical equilibrium. This is expected, as opacity sources that absorb strongly in the NUV will be depleted in the atmosphere at higher temperatures compared to chemical equilibrium if they rainout once condensation starts lower in the atmosphere. $\Delta$R$_{p,NUV-Red}$ is greater in the rainout models than in chemical equilibrium for the T=1887 and 2677~K models because TiO and VO also begin to rainout of the atmosphere at this temperature, decreasing the transit radius between 0.6 and 0.7 \microns{}, subsequently raising $\Delta$R$_{p,NUV-Red}$. $\Delta$R$_{p,NUV-Red}$ gradually decreases above 2677~K as H$^-$ opacity begins to raise the transit depths at red and IR wavelengths.

The behavior of $\Delta$R$_{p,Blue-Red}$ is somewhat more complicated. At 2000~K, there is a turnover from the index being dominated by Rayleigh scattering to being dominated by the gas opacities. For planets above 2000~K, we expect the gas opacities we describe here to result in a larger $\Delta$R$_{p,Blue-Red}$ than from Rayleigh scattering alone. Below 2000~K, however, we expect the blue transit depths caused by Rayleigh scattering to be larger than that caused by gas opacity.

\begin{figure*}
	\centering
	\gridline{\fig{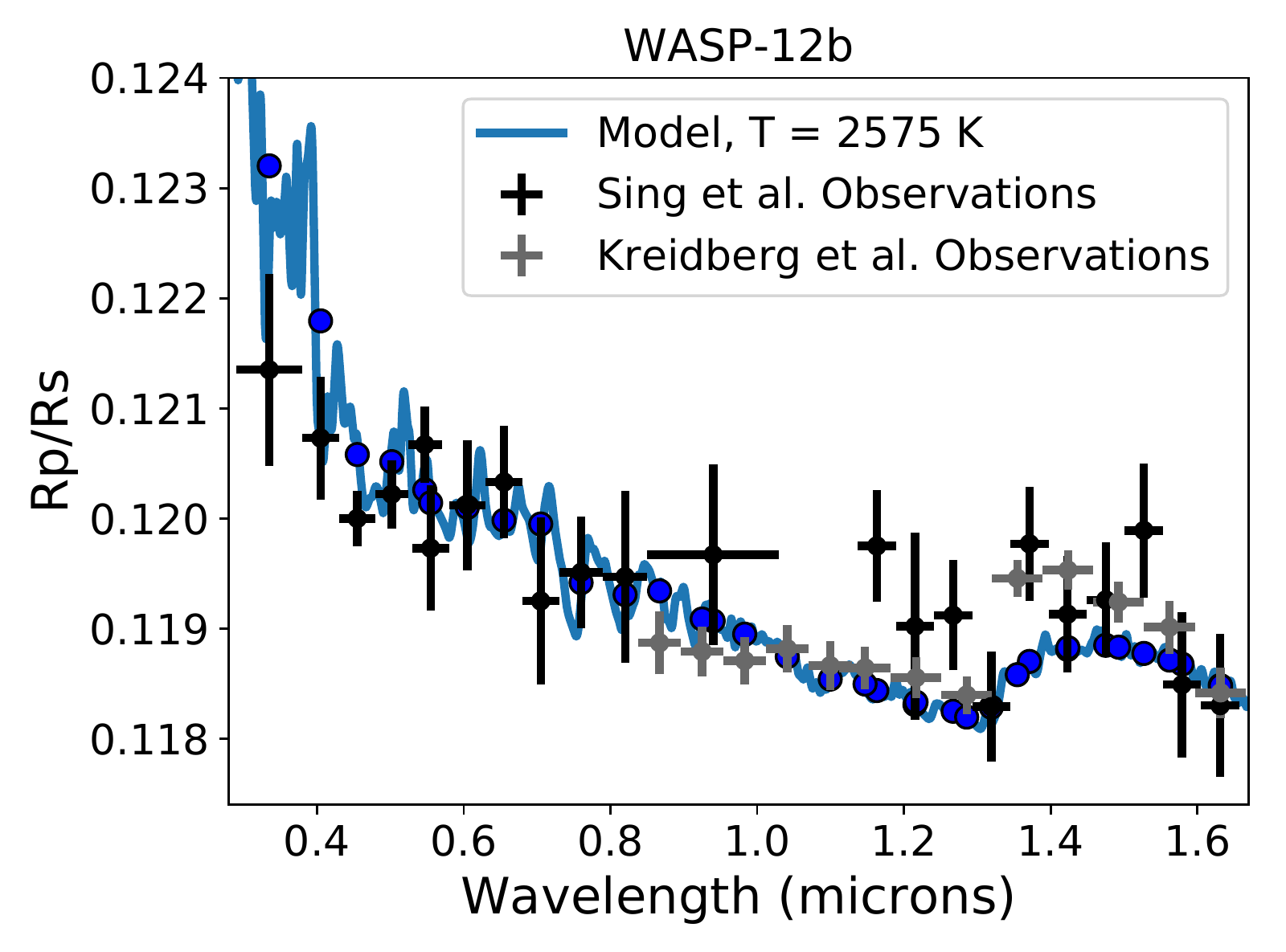}{0.5\textwidth}{}
		\fig{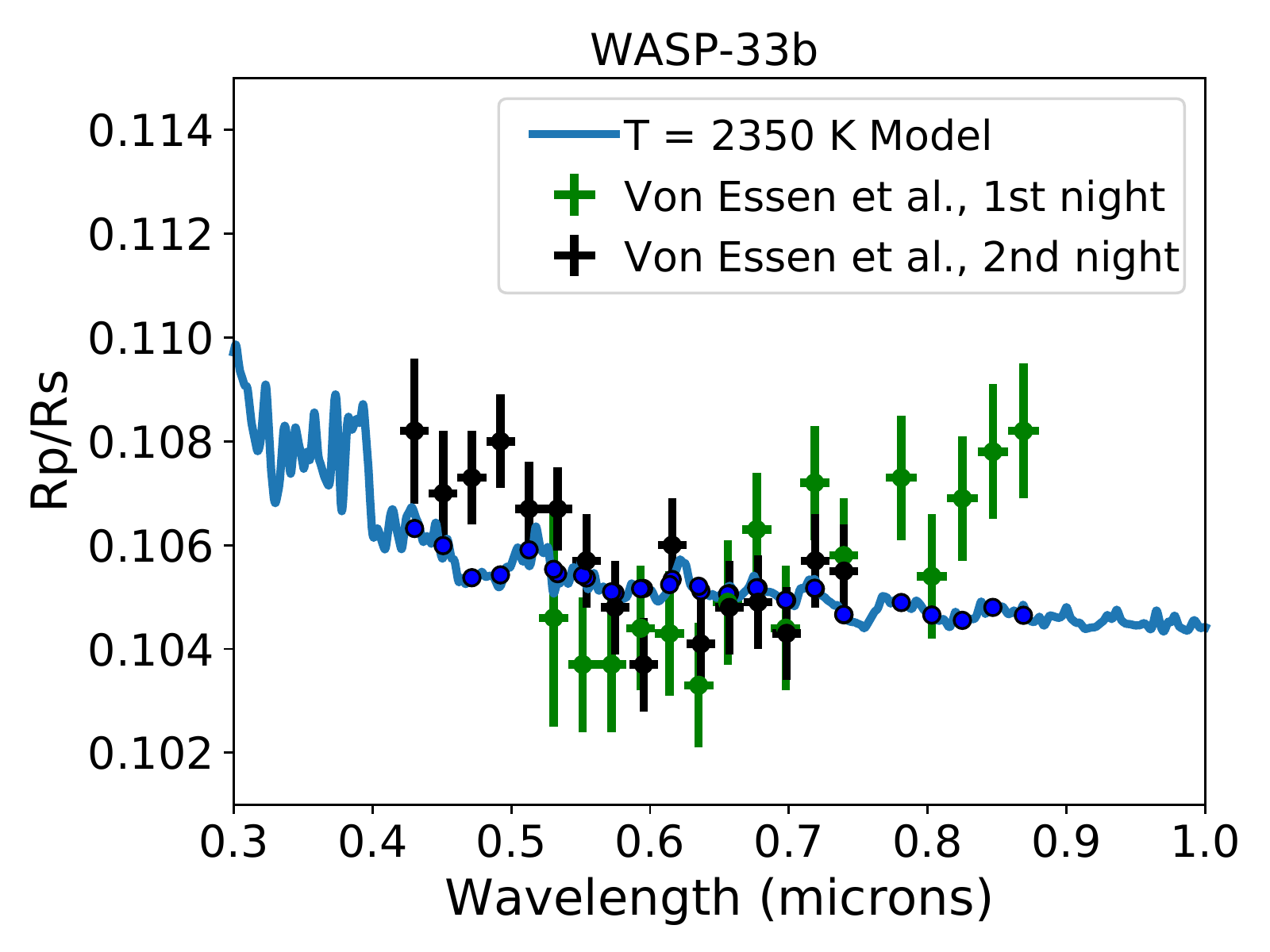}{0.5\textwidth}{}}
	\vspace{-25pt}
	\gridline{\fig{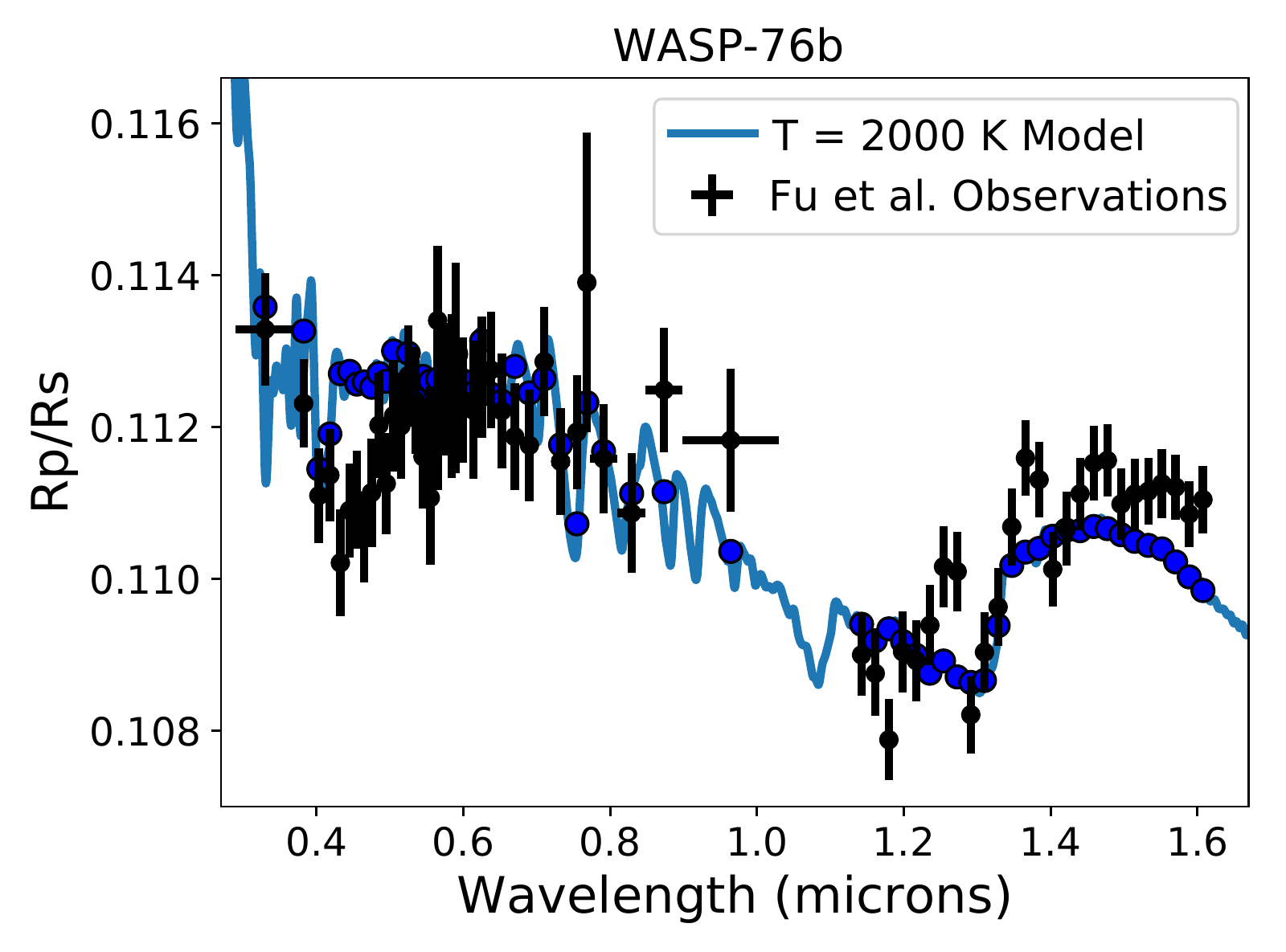}{0.5\textwidth}{}
		\fig{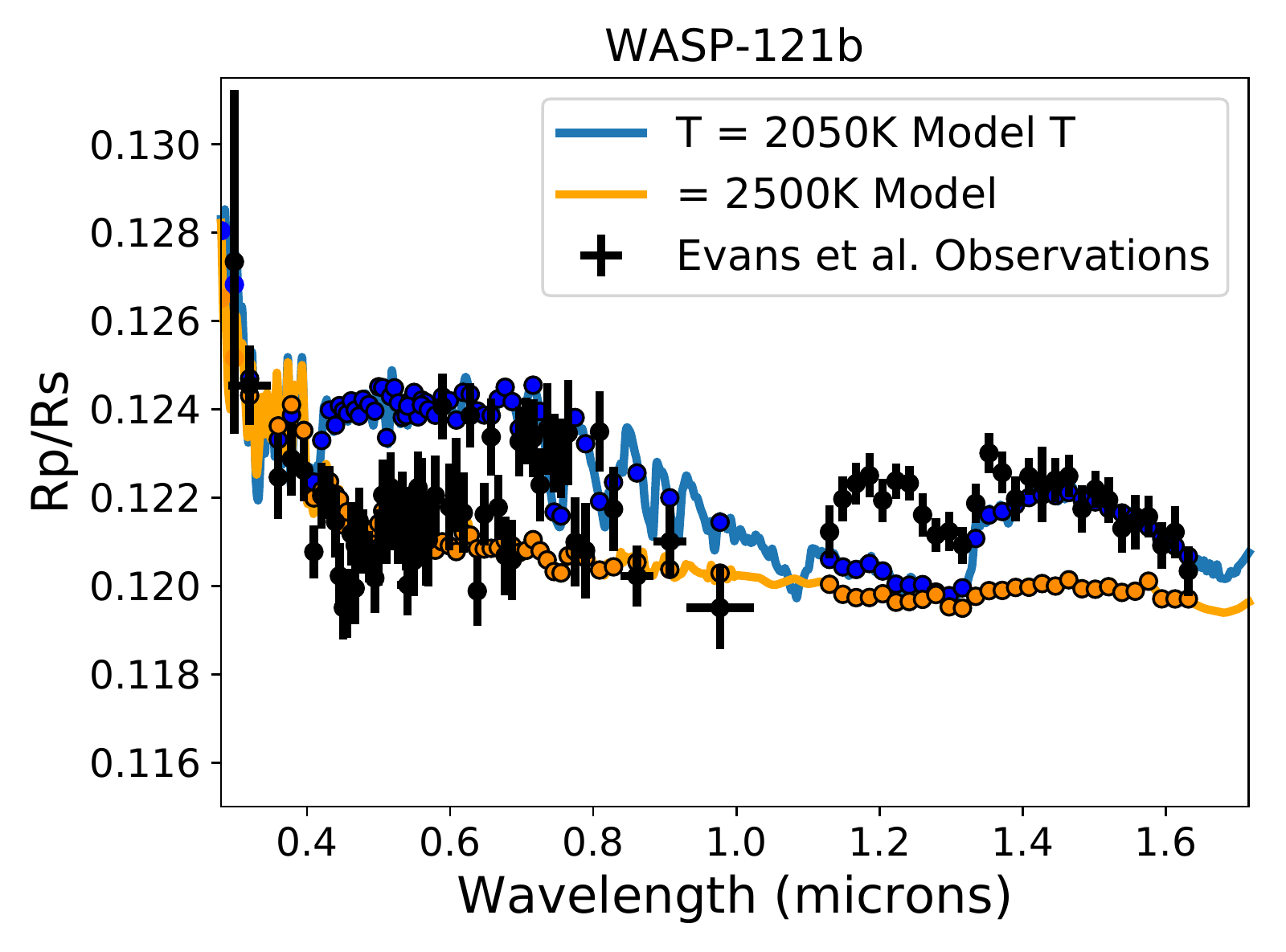}{0.5\textwidth}{}}
	
	\caption{Observed transit spectra of four ultra-hot Jupiters with wavelength coverage below 0.5 \microns{} compared to self-consistent models.\label{fig:obs}}
\end{figure*}

\subsection{Comparison to Previous Observations}
Only three hot Jupiters have reported observations between 0.2 and 0.3~\microns{}: WASP-12b \citep{fossati:2010}, HD~209458b \citep{vidal-madjar:2013}, and WASP-121b \citep{sing:2019}. WASP-12b and WASP-121b have $\Delta$R$_{p,NUV-Red}$ values of 66 and 25 planetary scale-heights, respectively, far above our model predictions. These wavelengths trace escaping (i.e., non-hydrostatic) gas, a process not accounted for in our hydrostatic models. For HD~209458b however, \cite{vidal-madjar:2013} and \cite{cubillos:2020} both find the observations consistent with Rayleigh scattering, but not precise enough to constrain its slope. Additional observations of these and other targets at short wavelengths, can constrain the behavior of the atmosphere (e.g., escaping versus hydrostatic), the metal content of hot Jupiter middle and upper atmospheres, and the effects of rainout chemistry as a function of planetary parameters.

Overplotted on Figure~\ref{fig:index} are the indices for a number of hot Jupiters from \cite{sing:2016} for which short-wavelength and NIR transit observations exist, allowing us to approximate the value of $\Delta$R$_{p,Blue-Red}$. These planets are plotted at their equilibrium temperatures. Interestingly, HD~209458b, whose optical spectrum is well-fit by our rainout model, is the only planet ${>2\sigma}$ below the maximum $\Delta$R$_{p,Blue-Red}$ suggested by Figure~\ref{fig:index}.

Also shown in Figure~\ref{fig:index} are the indices for the ultra-hot Jupiters WASP-12b, WASP-76b, and WASP-121b, whose full spectra are also shown in Figure~\ref{fig:obs}. These three hottest planets are each plotted at their best-fit terminator temperature, which is cooler than the planets' equilibrium temperature for WASP-12b and WASP-121b. For WASP-76b and WASP-121b, scattering is not indicated by the rest of their respective transit spectra and the short-wavelength in particular is well fit by our cloud- and haze-free models. Scattering by aerosols was, however, the original interpretation of the WASP-12b observations in \cite{sing:2013}. While we cannot unambiguously rule out scattering, the additional short-wavelength opacity discussed here provides a similarly well-fitting explanation of the data.

Our self-consistent models for these three planets have suitable fits, despite having no free parameters besides a small offset in R$_p$/R$_s$ and a rough fit in temperature (within a few hundred Kelvin). While the goodness-of-fit for the self-consistent model is comparable to that of the retrieved models for WASP-76b \citep{fu:2020}, adjusting additional parameters like metallicity or the temperature profile would likely improve the fit to each of the datasets. This is especially true for WASP-121b, for which retrievals prefer increased metallicity and a low temperature (compared to its equilibrium temperature). To fit the short-wavelength transit spectrum of WASP-121b with our self-consistent models, temperatures above WASP-121b's equilibrium temperature of $\sim${2350~K} fit the STIS/G430L observations, but do not fit the longer-wavelength STIS/G750L or WFC3/G141 data. Cooler self-consistent models can better match the STIS/G750L and WFC3/G141 data, but struggle to explain the STIS/G430L transit depths.

As is the case with WASP-76b mentioned above, our WASP-12b and WASP-121b model tends to overestimate the transit depth between about 0.43-0.5~\microns{}. This may be due to either condensation or rainout of some of the species discussed here, like Ca and Cr (See Figure~\ref{fig:contribution}). Condensation and rainout of Fe was recently reported on the nightside limb of WASP-76b \citep{ehrenreich:2020}.

In Figure~\ref{fig:obs}, we also plot transit spectrum of WASP-33b from 0.4 to 0.9 \microns{}. The spectrum stands out as rather poorly fit by our self-consistent models. Taken over the course of two nights at Gran Telescopio Canarias, these observations were originally interpreted as an indication of AlO opacity between 0.4 and 0.5 \microns{} \citep{vonessen:2018}. AlO would have to be highly out of equilibrium to explain the observed spectrum, as Al is preferentially in AlH or its atomic form at these temperatures.  We suggest further observations at these wavelengths for this valuable system to confirm the behavior of WASP-33b's short-wavelength transit spectrum. 

\section{Conclusions}\label{sec:conclude}

Inspired by the recent transit spectrum of ultra-hot Jupiter WASP-76b \citep{fu:2020}, we have modeled the short-wavelength transit spectrum of hot Jupiters. We show that large transit depths, especially for ultra-hot Jupiters, can be expected at NUV and blue wavelengths due to opacity from many species not considered at longer wavelengths, like Fe, Ti, Ni, and SiO. We emphasize that all the species discussed here are guaranteed to be present if conditions in the atmosphere are near chemical equilibrium and only efficient disequilibrium mechanisms, like cold-trapping and rainout, could preclude their imprint on short-wavelength transit spectra.

We defined two spectral indices to quantify the magnitude of the UV and blue-wavelength transit depths compared to red-wavelength transit depths as a function of temperature. We showed that the transit depth between 0.2 and 0.3 \microns{} is always larger than would be expected for Rayleigh scattering alone, except for in the coldest cases considered with rainout chemistry. At slightly longer wavelengths, between 0.3 and 0.4 \microns{}, there is a turnover compared to Rayleigh scattering at 2000~K, where Rayleigh scattering can dominate at lower temperatures, while gas-opacity will dominate at higher temperatures. The relationship between these indices and temperature changes if rainout chemistry is considered, providing a future path to characterizing processes occurring in the deeper atmosphere.

We recommend future observations at these wavelengths to identify the presence or absence of species. From space, this will require the use of HST/STIS or HST/WFC3/UVIS/G280, the latter of which shows great promise at characterizing the entire NUV-optical transit spectrum \citep{wakeford:2020} and could simultaneously measure the three spectral indices we have defined. Low-resolution, short-wavelength HST transit spectra would not only be able to detect opacity from metals, but would also be capable of detecting the effects of rainout if a variety of hot Jupiter targets are observed. Additionally, the Colorado Ultraviolet Transit Experiment (CUTE), slated to launch in 2020, will characterize the 2550-3300 \AA{} transmission spectrum of several hot and ultra-hot Jupiters at medium spectral-resolution with a 6-unit CubeSat \citep{fleming:2017}. By combining several transits together, CUTE has the potential to detect species like Mg I \& II at SNR$\sim$10-100. With HST/STIS/E230M and from the ground, high-resolution spectroscopy have already found success detecting and characterizing a plethora of species in the optical spectra of ultra-hot Jupiters \citep[e.g.,][]{sing:2019,hoeijmakers:2019,gibson:2020,ehrenreich:2020}. With more observations, trends between the presence and abundance of these species with planetary properties like equilibrium temperature, like those suggested in Figures~\ref*{fig:diffT} and \ref{fig:index}, can be identified.

\acknowledgments
This work is based on observations made with the NASA/ESA \emph{Hubble Space Telescope} obtained at the Space Telescope Science Institute, which is operated by the Association of Universities for Research in Astronomy, Inc. This research has made use of the NASA Astrophysics Data System and the NASA Exoplanet Archive, which is operated by the California Institute of Technology, under contract with the National Aeronautics and Space Administration under the Exoplanet Exploration Program.

\section*{}

\software{Matplotlib \citep{hunter:2007}, Numpy \citep{oliphant:2006,vanderwalt:2011}, Scipy \citep{virtanen:2020}, iPython \citep{perez:2007}}
\vspace{15pt}
\bibliographystyle{aasjournal}

\begin{thebibliography}{}
	\expandafter\ifx\csname natexlab\endcsname\relax\def\natexlab#1{#1}\fi
	\providecommand{\url}[1]{\href{#1}{#1}}
	\providecommand{\dodoi}[1]{doi:~\href{http://doi.org/#1}{\nolinkurl{#1}}}
	\providecommand{\doeprint}[1]{\href{http://ascl.net/#1}{\nolinkurl{http://ascl.net/#1}}}
	\providecommand{\doarXiv}[1]{\href{https://arxiv.org/abs/#1}{\nolinkurl{https://arxiv.org/abs/#1}}}
	
	\bibitem[{{Arcangeli} {et~al.}(2018){Arcangeli}, {D{\'e}sert}, {Line}, {Bean},
		{Parmentier}, {Stevenson}, {Kreidberg}, {Fortney}, {Mansfield}, \&
		{Showman}}]{arcangeli:2018}
	{Arcangeli}, J., {D{\'e}sert}, J.-M., {Line}, M.~R., {et~al.} 2018, \apjl, 855,
	L30, \dodoi{10.3847/2041-8213/aab272}
	
	\bibitem[{{Ballester} {et~al.}(2007){Ballester}, {Sing}, \&
		{Herbert}}]{ballester:2007}
	{Ballester}, G.~E., {Sing}, D.~K., \& {Herbert}, F. 2007, \nat, 445, 511,
	\dodoi{10.1038/nature05525}
	
	\bibitem[{{Barber} {et~al.}(2006){Barber}, {Tennyson}, {Harris}, \&
		{Tolchenov}}]{barber:2006}
	{Barber}, R.~J., {Tennyson}, J., {Harris}, G.~J., \& {Tolchenov}, R.~N. 2006,
	\mnras, 368, 1087, \dodoi{10.1111/j.1365-2966.2006.10184.x}
	
	\bibitem[{{Barman}(2007)}]{barman:2007}
	{Barman}, T. 2007, \apjl, 661, L191, \dodoi{10.1086/518736}
	
	\bibitem[{{Barman} {et~al.}(2001){Barman}, {Hauschildt}, \&
		{Allard}}]{barman:2001}
	{Barman}, T.~S., {Hauschildt}, P.~H., \& {Allard}, F. 2001, \apj, 556, 885,
	\dodoi{10.1086/321610}
	
	\bibitem[{{Bourrier} {et~al.}(2020){Bourrier}, {Ehrenreich}, {Lendl},
		{Cretignier}, {Allart}, {Dumusque}, {Cegla}, {Suarez-Mascareno},
		{Wyttenbach}, {Hoeijmakers}, {Melo}, {Kuntzer}, {Astudillo-Defru}, {Giles},
		{Heng}, {Kitzmann}, {Lavie}, {Lovis}, {Murgas}, {Nascimbeni}, {Pepe}, {Pino},
		{Segransan}, \& {Udry}}]{bourrier:2020}
	{Bourrier}, V., {Ehrenreich}, D., {Lendl}, M., {et~al.} 2020, arXiv e-prints,
	arXiv:2001.06836.
	\newblock \doarXiv{2001.06836}
	
	\bibitem[{{Cabot} {et~al.}(2020){Cabot}, {Madhusudhan}, {Welbanks}, {Piette},
		\& {Gandhi}}]{cabot:2020}
	{Cabot}, S. H.~C., {Madhusudhan}, N., {Welbanks}, L., {Piette}, A., \&
	{Gandhi}, S. 2020, arXiv e-prints, arXiv:2001.07196.
	\newblock \doarXiv{2001.07196}
	
	\bibitem[{{Casasayas-Barris} {et~al.}(2019){Casasayas-Barris}, {Pall{\'e}},
		{Yan}, {Chen}, {Kohl}, {Stangret}, {Parviainen}, {Helling}, {Watanabe},
		{Czesla}, {Fukui}, {Monta{\~n}{\'e}s-Rodr{\'\i}guez}, {Nagel}, {Narita},
		{Nortmann}, {Nowak}, {Schmitt}, \& {Zapatero Osorio}}]{casasayas:2019}
	{Casasayas-Barris}, N., {Pall{\'e}}, E., {Yan}, F., {et~al.} 2019, \aap, 628,
	A9, \dodoi{10.1051/0004-6361/201935623}
	
	\bibitem[{{Cubillos} {et~al.}(2020){Cubillos}, {Fossati}, {Koskinen}, {Young},
		{Salz}, {France}, {Sreejith}, \& {Haswell}}]{cubillos:2020}
	{Cubillos}, P.~E., {Fossati}, L., {Koskinen}, T., {et~al.} 2020, arXiv
	e-prints, arXiv:2001.03126.
	\newblock \doarXiv{2001.03126}
	
	\bibitem[{{Ehrenreich} {et~al.}(2020){Ehrenreich}, {Lovis}, {Allart}, {Rosa
			Zapatero Osorio}, {Pepe}, {Cristiani}, {Rebolo}, {Santos}, {Borsa},
		{Demangeon}, {Dumusque}, {Gonz{\'a}lez Hern{\'a}ndez}, {Casasayas-Barris},
		{S{\'e}gransan}, {Sousa}, {Abreu}, {Adibekyan}, {Affolter}, {Allende Prieto},
		{Alibert}, {Aliverti}, {Alves}, {Amate}, {Avila}, {Baldini}, {Bandy}, {Benz},
		{Bianco}, {Bolmont}, {Bouchy}, {Bourrier}, {Broeg}, {Cabral}, {Calderone},
		{Pall{\'e}}, {Cegla}, {Cirami}, {Coelho}, {Conconi}, {Coretti}, {Cumani},
		{Cupani}, {Dekker}, {Delabre}, {Deiries}, {D'Odorico}, {Di Marcantonio},
		{Figueira}, {Fragoso}, {Genolet}, {Genoni}, {G{\'e}nova Santos}, {Hara},
		{Hughes}, {Iwert}, {Kerber}, {Knudstrup}, {Land oni}, {Lavie}, {Lizon},
		{Lendl}, {Lo Curto}, {Maire}, {Manescau}, {Martins}, {M{\'e}gevand },
		{Mehner}, {Micela}, {Modigliani}, {Molaro}, {Monteiro}, {Monteiro},
		{Moschetti}, {M{\"u}ller}, {Nunes}, {Oggioni}, {Oliveira}, {Pariani},
		{Pasquini}, {Poretti}, {Rasilla}, {Redaelli}, {Riva}, {Santana Tschudi},
		{Santin}, {Santos}, {Segovia Milla}, {Seidel}, {Sosnowska}, {Sozzetti},
		{Span{\`o}}, {Su{\'a}rez Mascare{\~n}o}, {Tabernero}, {Tenegi}, {Udry},
		{Zanutta}, \& {Zerbi}}]{ehrenreich:2020}
	{Ehrenreich}, D., {Lovis}, C., {Allart}, R., {et~al.} 2020, arXiv e-prints,
	arXiv:2003.05528.
	\newblock \doarXiv{2003.05528}
	
	\bibitem[{{Evans} {et~al.}(2017){Evans}, {Sing}, {Kataria}, {Goyal}, {Nikolov},
		{Wakeford}, {Deming}, {Marley}, {Amundsen}, {Ballester}, {Barstow},
		{Ben-Jaffel}, {Bourrier}, {Buchhave}, {Cohen}, {Ehrenreich}, {Garc{\'{\i}}a
			Mu{\~n}oz}, {Henry}, {Knutson}, {Lavvas}, {Etangs}, {Lewis},
		{L{\'o}pez-Morales}, {Mandell}, {Sanz-Forcada}, {Tremblin}, \&
		{Lupu}}]{evans:2017}
	{Evans}, T.~M., {Sing}, D.~K., {Kataria}, T., {et~al.} 2017, \nat, 548, 58,
	\dodoi{10.1038/nature23266}
	
	\bibitem[{{Evans} {et~al.}(2018){Evans}, {Sing}, {Goyal}, {Nikolov}, {Marley},
		{Zahnle}, {Henry}, {Barstow}, {Alam}, {Sanz-Forcada}, {Kataria}, {Lewis},
		{Lavvas}, {Ballester}, {Ben-Jaffel}, {Blumenthal}, {Bourrier}, {Drummond},
		{Garc{\'\i}a Mu{\~n}oz}, {L{\'o}pez-Morales}, {Tremblin}, {Ehrenreich},
		{Wakeford}, {Buchhave}, {Lecavelier des Etangs}, {H{\'e}brard}, \&
		{Williamson}}]{evans:2018}
	{Evans}, T.~M., {Sing}, D.~K., {Goyal}, J.~M., {et~al.} 2018, \aj, 156, 283,
	\dodoi{10.3847/1538-3881/aaebff}
	
	\bibitem[{{Fleming} {et~al.}(2017){Fleming}, {France}, {Nell}, {Kohnert},
		{Pool}, {Egan}, {Fossati}, {Koskinen}, {Vidotto}, {Hoadley}, {Desert},
		{Beasley}, \& {Petit}}]{fleming:2017}
	{Fleming}, B.~T., {France}, K., {Nell}, N., {et~al.} 2017, in Society of
	Photo-Optical Instrumentation Engineers (SPIE) Conference Series, Vol. 10397,
	\procspie, 103971A, \dodoi{10.1117/12.2276138}
	
	\bibitem[{{Fossati} {et~al.}(2010){Fossati}, {Haswell}, {Froning}, {Hebb},
		{Holmes}, {Kolb}, {Helling}, {Carter}, {Wheatley}, {Collier Cameron},
		{Loeillet}, {Pollacco}, {Street}, {Stempels}, {Simpson}, {Udry}, {Joshi},
		{West}, {Skillen}, \& {Wilson}}]{fossati:2010}
	{Fossati}, L., {Haswell}, C.~A., {Froning}, C.~S., {et~al.} 2010, \apjl, 714,
	L222, \dodoi{10.1088/2041-8205/714/2/L222}
	
	\bibitem[{{Fu} {et~al.}(Submitted){Fu}, {Deming}, \& {Lothringer}}]{fu:2020}
	{Fu}, G., {Deming}, D., \& {Lothringer}, J.D., e.~a. Submitted, The
	Astronomical Journal
	
	\bibitem[{{Gandhi} \& {Madhusudhan}(2019)}]{gandhi:2019}
	{Gandhi}, S., \& {Madhusudhan}, N. 2019, \mnras, 485, 5817,
	\dodoi{10.1093/mnras/stz751}
	
	\bibitem[{{Gibson} {et~al.}(2020){Gibson}, {Merritt}, {Nugroho}, {Cubillos},
		{de Mooij}, {Mikal-Evans}, {Fossati}, {Lothringer}, {Nikolov}, {Sing},
		{Spake}, {Watson}, \& {Wilson}}]{gibson:2020}
	{Gibson}, N.~P., {Merritt}, S., {Nugroho}, S.~K., {et~al.} 2020, arXiv
	e-prints, arXiv:2001.06430.
	\newblock \doarXiv{2001.06430}
	
	\bibitem[{{Haswell} {et~al.}(2012){Haswell}, {Fossati}, {Ayres}, {France},
		{Froning}, {Holmes}, {Kolb}, {Busuttil}, {Street}, {Hebb}, {Collier Cameron},
		{Enoch}, {Burwitz}, {Rodriguez}, {West}, {Pollacco}, {Wheatley}, \&
		{Carter}}]{haswell:2012}
	{Haswell}, C.~A., {Fossati}, L., {Ayres}, T., {et~al.} 2012, \apj, 760, 79,
	\dodoi{10.1088/0004-637X/760/1/79}
	
	\bibitem[{{Hauschildt} {et~al.}(1999){Hauschildt}, {Allard}, \&
		{Baron}}]{hauschildt:1999}
	{Hauschildt}, P.~H., {Allard}, F., \& {Baron}, E. 1999, \apj, 512, 377,
	\dodoi{10.1086/306745}
	
	\bibitem[{{Hoeijmakers} {et~al.}(2019){Hoeijmakers}, {Ehrenreich}, {Kitzmann},
		{Allart}, {Grimm}, {Seidel}, {Wyttenbach}, {Pino}, {Nielsen}, {Fisher},
		{Rimmer}, {Bourrier}, {Cegla}, {Lavie}, {Lovis}, {Patzer}, {Stock}, {Pepe},
		\& {Heng}}]{hoeijmakers:2019}
	{Hoeijmakers}, H.~J., {Ehrenreich}, D., {Kitzmann}, D., {et~al.} 2019, \aap,
	627, A165, \dodoi{10.1051/0004-6361/201935089}
	
	\bibitem[{Hunter(2007)}]{hunter:2007}
	Hunter, J.~D. 2007, Computing in Science \& Engineering, 9, 90,
	\dodoi{10.1109/MCSE.2007.55}
	
	\bibitem[{{Kreidberg} {et~al.}(2015){Kreidberg}, {Line}, {Bean}, {Stevenson},
		{D{\'e}sert}, {Madhusudhan}, {Fortney}, {Barstow}, {Henry}, {Williamson}, \&
		{Showman}}]{kreidberg:2015}
	{Kreidberg}, L., {Line}, M.~R., {Bean}, J.~L., {et~al.} 2015, \apj, 814, 66,
	\dodoi{10.1088/0004-637X/814/1/66}
	
	\bibitem[{{Kurucz}(1993)}]{kurucz:1993}
	{Kurucz}, R. 1993, Diatomic Molecular Data for Opacity Calculations. Kurucz
	CD-ROM No. 15. Cambridge, 15
	
	\bibitem[{{Kurucz}(1994{\natexlab{a}})}]{kurucz:1994a}
	---. 1994{\natexlab{a}}, Atomic Data for Ca, 20
	
	\bibitem[{{Kurucz}(1994{\natexlab{b}})}]{kurucz:1994b}
	---. 1994{\natexlab{b}}, Atomic Data for Mn and Co. Kurucz CD-ROM No. 21.
	Cambridge, 21
	
	\bibitem[{{Kurucz}(1994{\natexlab{c}})}]{kurucz:1994c}
	---. 1994{\natexlab{c}}, Atomic Data for Fe and Ni. Kurucz CD-ROM No. 22.
	Cambridge, 22
	
	\bibitem[{{Lecavelier Des Etangs} {et~al.}(2008){Lecavelier Des Etangs},
		{Pont}, {Vidal-Madjar}, \& {Sing}}]{etangs:2008}
	{Lecavelier Des Etangs}, A., {Pont}, F., {Vidal-Madjar}, A., \& {Sing}, D.
	2008, \aap, 481, L83, \dodoi{10.1051/0004-6361:200809388}
	
	\bibitem[{{Lothringer} \& {Barman}(2019)}]{lothringer:2019}
	{Lothringer}, J.~D., \& {Barman}, T. 2019, \apj, 876, 69,
	\dodoi{10.3847/1538-4357/ab1485}
	
	\bibitem[{{Lothringer} {et~al.}(2018){Lothringer}, {Barman}, \&
		{Koskinen}}]{lothringer:2018b}
	{Lothringer}, J.~D., {Barman}, T., \& {Koskinen}, T. 2018, \apj, 866, 27,
	\dodoi{10.3847/1538-4357/aadd9e}
	
	\bibitem[{{McCullough} {et~al.}(2014){McCullough}, {Crouzet}, {Deming}, \&
		{Madhusudhan}}]{mccullough:2014}
	{McCullough}, P.~R., {Crouzet}, N., {Deming}, D., \& {Madhusudhan}, N. 2014,
	\apj, 791, 55, \dodoi{10.1088/0004-637X/791/1/55}
	
	\bibitem[{Oliphant(2006)}]{oliphant:2006}
	Oliphant, T.~E. 2006, A guide to NumPy, Vol.~1 (Trelgol Publishing USA)
	
	\bibitem[{{Parmentier} {et~al.}(2018){Parmentier}, {Line}, {Bean}, {Mansfield},
		{Kreidberg}, {Lupu}, {Visscher}, {D{\'e}sert}, {Fortney}, {Deleuil},
		{Arcangeli}, {Showman}, \& {Marley}}]{parmentier:2018}
	{Parmentier}, V., {Line}, M.~R., {Bean}, J.~L., {et~al.} 2018, \aap, 617, A110,
	\dodoi{10.1051/0004-6361/201833059}
	
	\bibitem[{{Pont} {et~al.}(2008){Pont}, {Knutson}, {Gilliland}, {Moutou}, \&
		{Charbonneau}}]{pont:2008}
	{Pont}, F., {Knutson}, H., {Gilliland}, R.~L., {Moutou}, C., \& {Charbonneau},
	D. 2008, \mnras, 385, 109, \dodoi{10.1111/j.1365-2966.2008.12852.x}
	
	\bibitem[{{Pont} {et~al.}(2013){Pont}, {Sing}, {Gibson}, {Aigrain}, {Henry}, \&
		{Husnoo}}]{pont:2013}
	{Pont}, F., {Sing}, D.~K., {Gibson}, N.~P., {et~al.} 2013, ArXiv e-prints.
	\newblock \doarXiv{1210.4163}
	
	\bibitem[{PŽrez \& Granger(2007)}]{perez:2007}
	PŽrez, F., \& Granger, B.~E. 2007, Computing in Science \& Engineering, 9, 21,
	\dodoi{10.1109/MCSE.2007.53}
	
	\bibitem[{{Rackham} {et~al.}(2018){Rackham}, {Apai}, \&
		{Giampapa}}]{rackham:2018}
	{Rackham}, B.~V., {Apai}, D., \& {Giampapa}, M.~S. 2018, \apj, 853, 122,
	\dodoi{10.3847/1538-4357/aaa08c}
	
	\bibitem[{{Salz} {et~al.}(2019){Salz}, {Schneider}, {Fossati}, {Czesla},
		{France}, \& {Schmitt}}]{salz:2019}
	{Salz}, M., {Schneider}, P.~C., {Fossati}, L., {et~al.} 2019, \aap, 623, A57,
	\dodoi{10.1051/0004-6361/201732419}
	
	\bibitem[{{Schwenke}(1998)}]{schwenke:1998}
	{Schwenke}, D.~W. 1998, Faraday Discussions, 109, 321, \dodoi{10.1039/a800070k}
	
	\bibitem[{{Sing} {et~al.}(2008){Sing}, {Vidal-Madjar}, {D{\'e}sert},
		{Lecavelier des Etangs}, \& {Ballester}}]{sing:2008b}
	{Sing}, D.~K., {Vidal-Madjar}, A., {D{\'e}sert}, J.-M., {Lecavelier des
		Etangs}, A., \& {Ballester}, G. 2008, \apj, 686, 658, \dodoi{10.1086/590075}
	
	\bibitem[{{Sing} {et~al.}(2013){Sing}, {Lecavelier des Etangs}, {Fortney},
		{Burrows}, {Pont}, {Wakeford}, {Ballester}, {Nikolov}, {Henry}, {Aigrain},
		{Deming}, {Evans}, {Gibson}, {Huitson}, {Knutson}, {Showman}, {Vidal-Madjar},
		{Wilson}, {Williamson}, \& {Zahnle}}]{sing:2013}
	{Sing}, D.~K., {Lecavelier des Etangs}, A., {Fortney}, J.~J., {et~al.} 2013,
	\mnras, 436, 2956, \dodoi{10.1093/mnras/stt1782}
	
	\bibitem[{{Sing} {et~al.}(2016){Sing}, {Fortney}, {Nikolov}, {Wakeford},
		{Kataria}, {Evans}, {Aigrain}, {Ballester}, {Burrows}, {Deming},
		{D{\'e}sert}, {Gibson}, {Henry}, {Huitson}, {Knutson}, {Etangs}, {Pont},
		{Showman}, {Vidal-Madjar}, {Williamson}, \& {Wilson}}]{sing:2016}
	{Sing}, D.~K., {Fortney}, J.~J., {Nikolov}, N., {et~al.} 2016, \nat, 529, 59,
	\dodoi{10.1038/nature16068}
	
	\bibitem[{{Sing} {et~al.}(2019){Sing}, {Lavvas}, {Ballester}, {Lecavelier des
			Etangs}, {Marley}, {Nikolov}, {Ben-Jaffel}, {Bourrier}, {Buchhave}, {Deming},
		{Ehrenreich}, {Mikal-Evans}, {Kataria}, {Lewis}, {L{\'o}pez-Morales},
		{Garc{\'\i}a Mu{\~n}oz}, {Henry}, {Sanz-Forcada}, {Spake}, {Wakeford}, \&
		{The PanCET collaboration}}]{sing:2019}
	{Sing}, D.~K., {Lavvas}, P., {Ballester}, G.~E., {et~al.} 2019, \aj, 158, 91,
	\dodoi{10.3847/1538-3881/ab2986}
	
	\bibitem[{{van der Walt} {et~al.}(2011){van der Walt}, {Colbert}, \&
		{Varoquaux}}]{vanderwalt:2011}
	{van der Walt}, S., {Colbert}, S.~C., \& {Varoquaux}, G. 2011, Computing in
	Science and Engineering, 13, 22, \dodoi{10.1109/MCSE.2011.37}
	
	\bibitem[{{Vidal-Madjar} {et~al.}(2013){Vidal-Madjar}, {Huitson}, {Bourrier},
		{D{\'e}sert}, {Ballester}, {Lecavelier des Etangs}, {Sing}, {Ehrenreich},
		{Ferlet}, {H{\'e}brard}, \& {McConnell}}]{vidal-madjar:2013}
	{Vidal-Madjar}, A., {Huitson}, C.~M., {Bourrier}, V., {et~al.} 2013, \aap, 560,
	A54, \dodoi{10.1051/0004-6361/201322234}
	
	\bibitem[{{Virtanen} {et~al.}(2020){Virtanen}, {Gommers}, {Oliphant},
		{Haberland}, {Reddy}, {Cournapeau}, {Burovski}, {Peterson}, {Weckesser},
		{Bright}, {van der Walt}, {Brett}, {Wilson}, {Jarrod Millman}, {Mayorov},
		{Nelson}, {Jones}, {Kern}, {Larson}, {Carey}, {Polat}, {Feng}, {Moore}, {Vand
			erPlas}, {Laxalde}, {Perktold}, {Cimrman}, {Henriksen}, {Quintero}, {Harris},
		{Archibald}, {Ribeiro}, {Pedregosa}, {van Mulbregt}, \&
		{Contributors}}]{virtanen:2020}
	{Virtanen}, P., {Gommers}, R., {Oliphant}, T.~E., {et~al.} 2020, Nature
	Methods, \dodoi{https://doi.org/10.1038/s41592-019-0686-2}
	
	\bibitem[{{von Essen} {et~al.}(2019){von Essen}, {Mallonn}, {Welbanks},
		{Madhusudhan}, {Pinhas}, {Bouy}, \& {Weis Hansen}}]{vonessen:2018}
	{von Essen}, C., {Mallonn}, M., {Welbanks}, L., {et~al.} 2019, \aap, 622, A71,
	\dodoi{10.1051/0004-6361/201833837}
	
	\bibitem[{{Wakeford} {et~al.}(2020){Wakeford}, {Sing}, {Stevenson}, {Lewis},
		{Pirzkal}, {Wilson}, {Goyal}, {Kataria}, {Mikal-Evans}, {Nikolov}, \&
		{Spake}}]{wakeford:2020}
	{Wakeford}, H.~R., {Sing}, D.~K., {Stevenson}, K.~B., {et~al.} 2020, arXiv
	e-prints, arXiv:2003.00536.
	\newblock \doarXiv{2003.00536}
	
	\bibitem[{{Yan} \& {Henning}(2018)}]{yan:2018}
	{Yan}, F., \& {Henning}, T. 2018, Nature Astronomy, 2, 714,
	\dodoi{10.1038/s41550-018-0503-3}
	
	\bibitem[{{Zahnle} {et~al.}(2009){Zahnle}, {Marley}, {Freedman}, {Lodders}, \&
		{Fortney}}]{zahnle:2009b}
	{Zahnle}, K., {Marley}, M.~S., {Freedman}, R.~S., {Lodders}, K., \& {Fortney},
	J.~J. 2009, \apjl, 701, L20, \dodoi{10.1088/0004-637X/701/1/L20}
	
	\bibitem[{{Zhang} {et~al.}(2019){Zhang}, {Chachan}, {Kempton}, \&
		{Knutson}}]{zhang:2019}
	{Zhang}, M., {Chachan}, Y., {Kempton}, E. M.~R., \& {Knutson}, H.~A. 2019,
	\pasp, 131, 034501, \dodoi{10.1088/1538-3873/aaf5ad}
	
\end{thebibliography}

\end{document}